\documentclass{svmult}

\usepackage{makeidx}         
\usepackage{graphicx}        
\usepackage[tight,footnotesize]{subfigure}
\usepackage{multicol}        
\usepackage[bottom]{footmisc}

\makeindex             

\begin{document}
%
\title*{Quantitative Verification of a Force-based Model for Pedestrian 
Dynamics}
\titlerunning{Quantitative Verification of a force-based Model}
\author{Mohcine Chraibi\inst{1}\and
Armin Seyfried\inst{1}\and
Andreas Schadschneider\inst{2}\and
Wolfgang Mackens\inst{3}}

\institute{J\"ulich Supercomputing Centre, Forschungszentrum J\"ulich, 
52425 J\"ulich, Germany.
\texttt{m.chraibi@fz-juelich.de, a.seyfried@fz-juelich.de}
\and Institute for Theoretical Physics, Universit\"at zu K\"oln, D-50937
K\"oln, Germany.
 \texttt{as@thp.uni-koeln.de}
\and Hamburg University of Technology, 21071 Hamburg, Germany.
   \texttt{mackens@tuhh.de}
}
\authorrunning{Chraibi, Seyfried, Schadschneider, Mackens}
\maketitle

\begin{abstract}
  This paper introduces a spatially continuous force-based model for
  simulating pedestrian dynamics.  The main intention of this work is
  the quantitative description of pedestrian movement through
  bottlenecks and in corridors. Measurements of flow and density at
  bottlenecks will be presented and compared with empirical data.
  Furthermore the fundamental diagram for the movement in a corridor
  is reproduced.  The results of the proposed model show a good
  agreement with empirical data.
\end{abstract}

\section{Introduction}
One application of pedestrian dynamics is the enhancement of the
safety of people in complex buildings and in big mass
events e.g., sporting events, religious pilgrimages, etc. where there is
a risk of disaster. Thanks to computer simulations, it is possible
to forecast the emergency egress and optimise the evacuation of large
crowds. Another aspect of pedestrian dynamics is the comfort of
passengers in pedestrian facilities e.g., airports, railway stations,
shopping malls, etc.  Those facilities have to be designed in a way to
ensure minimal travel times and maximal capacities.  For these
applications, robust and quantitatively validated models are
necessary.

A wide spectrum of models have been designed to simulate pedestrian
dynamics. Generally those models can be classified into macroscopic
and microscopic models. In macroscopic models the system is described
by mean values of characteristics of pedestrian streams e.g., density
and flow, whereas microscopic models consider the movement of
individual persons separately. Microscopic models can be subdivided into
several classes e.g., rule-based and force-based models.  For a
detailed discussion we refer to~\cite{Schadschneider2009a}. In this
work we focus on spatially continuous force-based models.

Force-based models take Newton's second law of dynamics as a guiding
principle. Thus, the movement of each pedestrian is defined by:  
 \begin{equation}
\overrightarrow{F_{i}} = \sum_{j\neq i}^{\tilde N}  
\overrightarrow{F_{ij}^{\rm rep}} + \sum_{B} 
\overrightarrow{F_{iB}^{\rm rep}} + \overrightarrow{F_{i}^{\rm drv}} 
= m_i\overrightarrow{a_i}, 
\label{eq:maineq}
\end{equation}
where $\overrightarrow{F_{ij}^{\rm rep}}$ denotes the repulsive force
from pedestrian $j$ acting on pedestrian $i$,
$\overrightarrow{F_{iB}^{\rm rep}}$ is the repulsive force emerging from
borders and $\overrightarrow{F_{i}^{\rm drv}}$ is a driving force. $m_i$
is a constant with dimensions of mass and $\tilde N$ the number of
neighbouring pedestrians.  Repulsive forces model the
collision-avoidance performed by pedestrians. Whereas the driving
force models the intention of a pedestrian to move to some
destination.
The set of equations~(\ref{eq:maineq}) for all pedestrians
results in a high-dimensional system of second order ordinary differential
equations. The time evolution of the positions and velocities 
of all pedestrians is obtained by numerical integration.

Most force-based models describe the movement of pedestrians
qualitatively well. Collective phenomena like lane
formations~\cite{Helbing1995,Helbing2004,Yu2005}, oscillations at
bottlenecks~\cite{Helbing1995,Helbing2004}, the ``faster-is-slower''
effect \cite{Lakoba2005,Parisi2007}, clogging at exit
doors~\cite{Helbing2004,Yu2005} etc. are reproduced.  These
achievements indicate that these models are promising candidates.
However, a qualitative description is not sufficient if reliable
statements about critical processes, e.g., emergency egress, are
requested. Moreover, implementations of models do not
rely on one sole approach. Especially in high density situations
simple numerical treatment has to be supplemented by additional
techniques to obtain reasonable results. Examples are restrictions on
state variables and sometimes even totally different procedures
replacing the above equations of motion~(\ref{eq:maineq}) to avoid
partial and total overlapping among pedestrians
\cite{Lakoba2005,Yu2005} or negative and high velocities
\cite{Helbing1995}.

We address the possibility of describing reasonably and in a
quantitative manner the movement of pedestrians, with a modelling
approach as simple as possible. For a systematic verification of our
model we measure the fundamental diagram, the flow through bottlenecks
and the density inside and in front of the entrance of a bottleneck.
In the next section, we propose such a model which is solely based on the
equation of motion~(\ref{eq:maineq}). Furthermore the model
incorporates free parameters which allow calibration to fit
quantitative data.

\section{Definition of the model}

\begin{figure}
\begin{center}
\includegraphics[scale=0.70]{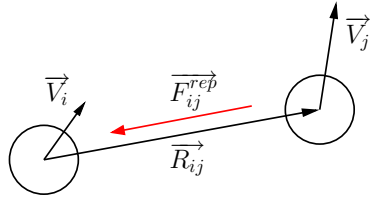}
\caption{The direction of the repulsive force pedestrian $j$\, acting on
 pedestrian $i$.}
\label{fig:fuss}
\end{center}
\end{figure}

Our model is based on the Centrifugal Force Model (CFM)~\cite{Yu2005}.
The CFM takes into account the distance between pedestrians as well as
their relative velocities. Pedestrians are modelled as circles with constant diameter.  Their movement is a direct result of
superposition of repulsive and driving forces acting on the centre of
each pedestrian.  Repulsive forces acting on pedestrian $i$ from other
pedestrians in their neighbourhood and eventually from walls, stairs,
etc. to prevent collisions and overlapping (Fig.~\ref{fig:fuss}).
The driving force, however, adds a positive term to the resulting
force, to enable movement of pedestrian $i$ in a certain direction
with a given desired speed $\parallel\overrightarrow{V_{i}^0}\parallel$.  The
mathematical expression of the driving force as introduced initially
in~\cite{Helbing1995} is used:
\begin{equation}
\overrightarrow{F_{i}^{\rm drv}} = m_i \frac{\overrightarrow{V_{i}^0} -
 \overrightarrow{V_{i}}} {{\tau}},
\label{eq:fdrv}
\end{equation}
with a time constant ${\tau}$. 

The definition of the repulsive force in the CFM expresses several principles. 
First, the force between two pedestrians decreases with increasing 
distance. In the CFM it is inversely proportional to their distance.
Given the position of two pedestrians $i$ and $j$, the direction
vector between their centers is defined as:
 \begin{equation}
 \overrightarrow{R_{ij}} = \overrightarrow{R_j} -\overrightarrow{R_i}
,\;\;\;\; 
\overrightarrow{e_{ij}} = \frac{\overrightarrow{R_{ij}}}{\parallel 
\overrightarrow{R_{ij}} \parallel}\,. 
\label{eq:R}
\end{equation} 
Furthermore, the repulsive force takes into account the relative
velocity between pedestrian $i$ and pedestrian $j$. The following
special definition provides that slower pedestrians are not affected
by the presence of faster pedestrians in front of them:
\begin{equation}
 V_{ij} = \frac{1}{2}[(\overrightarrow{V_i}-\overrightarrow{V_j})\cdot
\overrightarrow{e_{ij}} + |(\overrightarrow{V_i}-\overrightarrow{V_j})
\cdot\overrightarrow{e_{ij}}|].
\label{eq:relv}
\end{equation} 
As in general pedestrians  react only to obstacles and pedestrians that are
within their perception, the reaction field of the repulsive force is
reduced to the angle of vision of each pedestrian ($180^\circ$), by
introducing the coefficient: 
\begin{equation}
K_{ij} = \frac{1}{2}\frac{\overrightarrow{V_i}\cdot\overrightarrow{e_{ij}}
+ \mid \overrightarrow{V_i}\cdot\overrightarrow{e_{ij}} \mid} 
{\parallel \overrightarrow{V_i}\parallel}.
\label{eq:K}
 \end{equation}
With the definitions in Eqs.~(\ref{eq:R}), (\ref{eq:relv}) and
(\ref{eq:K}), the repulsive force between two pedestrians is formulated as:
  \begin{equation}
 \overrightarrow{F_{ij}^{\rm rep}} = -m_i K_{ij} \frac{V_{ij}^2}{\parallel
\overrightarrow{R_{ij}}\parallel}\overrightarrow{e_{ij}}\,.
\label{eq:CFMfrep}
\end{equation}
In~\cite{Chraibi2009a} it was shown that the introduction of a
``collision detection technique'' (CDT), see~\cite{Yu2005} for the
definition, is necessary to mitigate overlapping among pedestrians.

In the following, we will discuss why volume exclusion is not
guaranteed by Eq.~(\ref{eq:CFMfrep}) and meanwhile introduce our
modifications of the repulsive force. 
Due to the quotient in Eq.~(\ref{eq:CFMfrep}) when the distance is
small, low relative velocities 
lead to an unacceptably small
force. Consequently, partial or total overlapping are not prevented. Introducing the intended speed
in the numerator of the repulsive force eliminates this side-effect.
Furthermore, the modified repulsive force and driving
force~(\ref{eq:fdrv}) compensate at low velocities, which damps
oscillations.

Since faster pedestrians require more space than slower pedestrians,
due to increasing step sizes~\cite{Seyfried2006}, the diameter of
pedestrian $i$ depends linearly on its velocity:
\begin{equation}
D_i= d_a + d_b\parallel\overrightarrow{V_i}\parallel,
\label{eq:diam}
\end{equation}
with free parameters $d_a$\, and $d_b$. We define the distance between
pedestrian $i$ and pedestrian $j$ as:
\begin{equation}
 {\rm dist}_{ij}=\parallel
  \overrightarrow{R_{ij}}\parallel-\frac{1}{2}(D_i(\parallel
\overrightarrow{V_i}\parallel)+D_j(\parallel\overrightarrow{V_j}\parallel)).
\label{eq:dist-eff}
\end{equation}

By taking these aspects into account, the definition of the modified
repulsive force reads
\begin{equation}
\overrightarrow{F_{ij}^{\rm rep}}=-m_i K_{ij}\frac{(\nu \parallel
\overrightarrow{V_i^0}\parallel + V_{ij})^2}{{\rm dist}_{ij}}
\overrightarrow{e_{ij}},
\label{eq:frep}
\end{equation}
where $\nu$ is a parameter which adjusts the strength of the force.
Due to these changes we can do without the extra CDT which dominates the dynamics
in \cite{Yu2005} in case of formation of dense crowds.

\begin{figure}
 \begin{center}
  \subfigure{\label{fig:interp}}
  \includegraphics[scale=0.47]{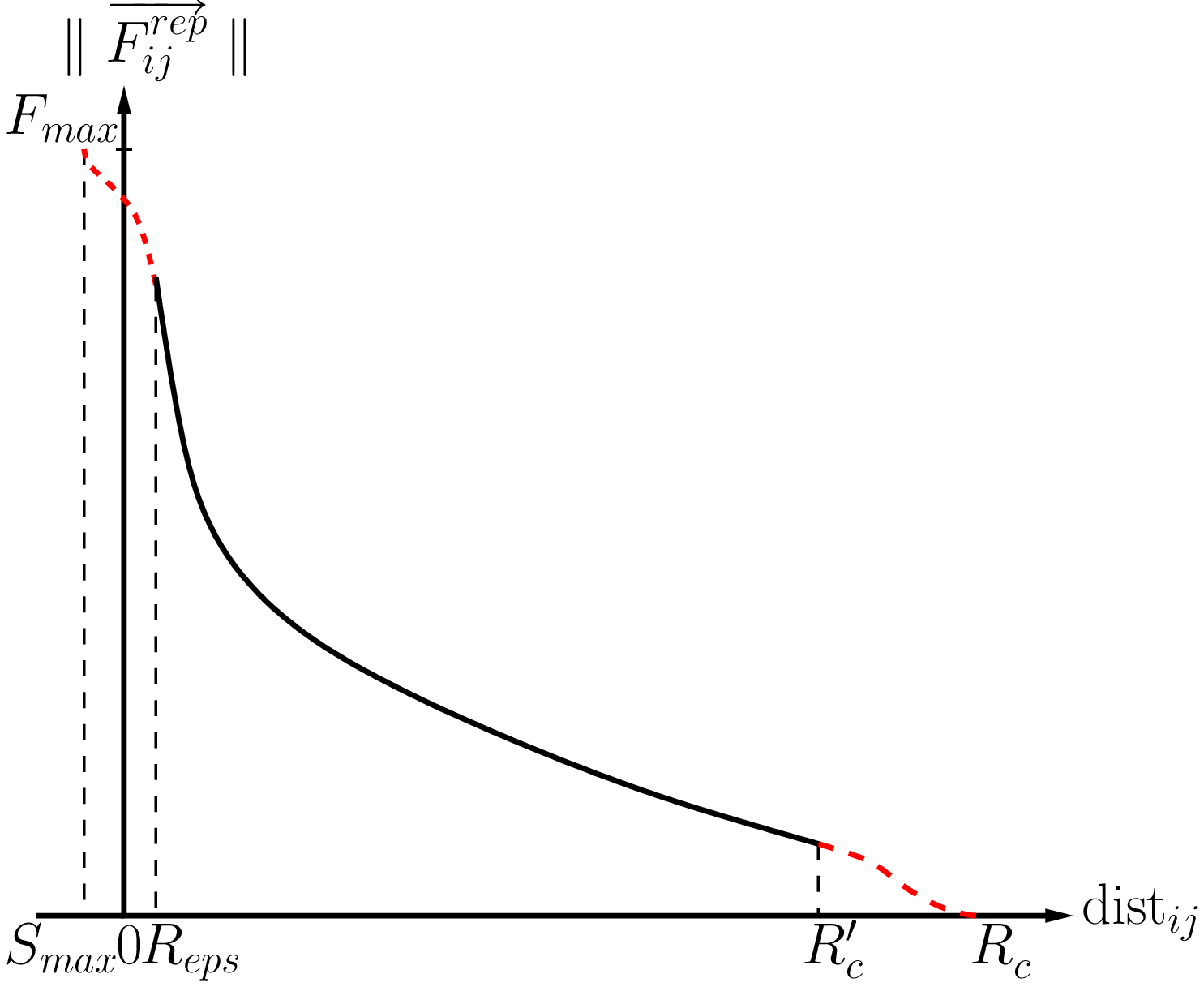}
  \subfigure{\label{fig:dir}}
  \includegraphics[scale=0.47]{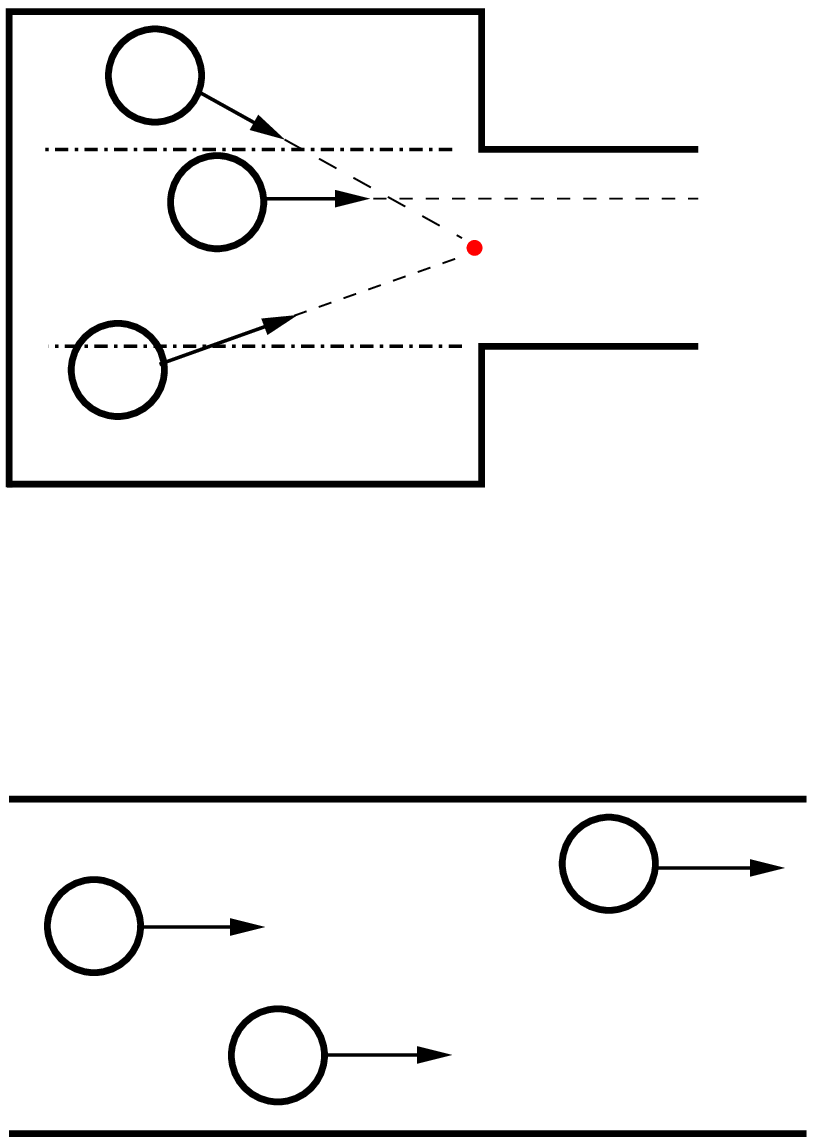}
  \caption{Left: The interpolation of the repulsive force between pedestrians
  $i$\, and $j$. Right: Direction of pedestrians in corridors and bottlenecks.}
  \label{fig:interp-dir}
 \end{center}
\end{figure}

The repulsive force between two pedestrians $i$\, and $j$ is infinite
at contact and decreasing with increasing distance between $i$ and
$j$. Since the repulsive force as defined in Eq.~(\ref{eq:frep}) does
not vanish, the summation over all other pedestrians leads to a
complexity of $O(N^2)$. To deal with this problem and to consider a
limited range of pedestrian interaction only the influence of
neighbouring pedestrians is taken into account.  Two pedestrians are
said to be neighbours if their distance is within a certain cut-off
radius $R_c=2.5\;\mbox{m}$. To guarantee robust numerical integration
a two-sided Hermite-interpolation of the repulsive force is
implemented (see Fig.~\ref{fig:interp-dir}).
The interpolation guarantees that for each pair $i$, $j$ with a distance in the interval $[{R^\prime}_{c}, R_c]$ the norm
of the repulsive force between them decreases smoothly to zero.
${R^\prime}_{c}$ is set to $R_c-0.1\; \mbox{m} $.
For distances in the interval $[S_{\rm max}, R_{\rm eps}]$ the interpolation
avoids an increase of the force to infinity, to reach a maximum value
of $F_{\rm max}=1000\;\mbox{N}$.  $R_{\rm eps}$ is set to $0.1\; \mbox{m}$
and $S_{\rm max}$ to $-5\;\mbox{m}$.

The desired direction of a pedestrian is set to be parallel to the
walls of the corridor. In the bottleneck case it is set towards the
centre of the entrance to the bottleneck if the pedestrian is outside
the range of the bottleneck. That is if he can not ``see'' the exit of
the bottleneck. Otherwise, the desired direction is chosen parallel to
the length of the bottleneck (Fig.~\ref{fig:interp-dir}).

\section{Simulation results}

The initial value problem~(\ref{eq:maineq}) was solved using an Euler
scheme with fixed-step size $\Delta t =0.01\;$s. The desired speeds of
pedestrians are Gaussian distributed with mean $\mu=1.34\;\mbox{m/s}$ and
standard deviation $\sigma=0.26\;\mbox{m/s}$. The constant ${\tau}$ in
Eq.~(\ref{eq:fdrv}) is set to $0.5\;\, \mbox{s}$. For simplicity, the
mass, $m_i$ is set to unity. Several parameter values were tested.
The free parameters in Eqs.~(\ref{eq:frep}) and~(\ref{eq:diam}) are
set to $\nu = 0.2$, $d_a=0.3\mbox{ m}\;\, \mbox{and}\;\; d_b=0.2\mbox{
  s} $. With this parameter set the results of the simulations are in
good agreement with empirical data.

To verify the ability of the model to reproduce the fundamental diagram,
measurements in corridors of different widths were performed. The length
of the corridor is $20\;\mbox{m}$ and its width is $2\;\mbox{m}$.
\begin{figure}[htp]
  \begin{center}
   \subfigure{\label{fig:fd2d}
   \includegraphics[scale=0.32, angle=270]{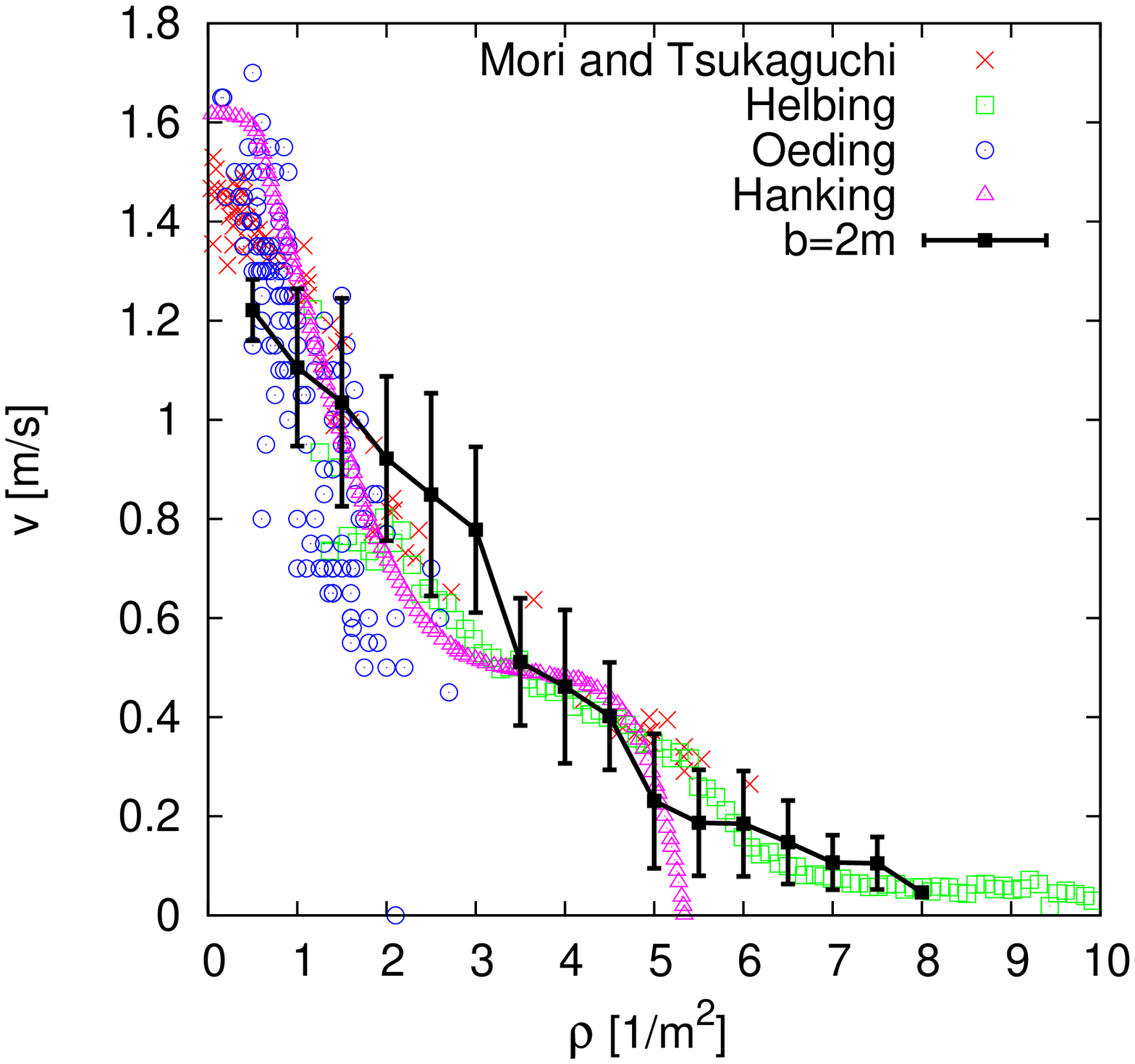}}
 \hfill
\subfigure{\label{fig:flowa}
 \includegraphics[scale=0.31, angle=270]{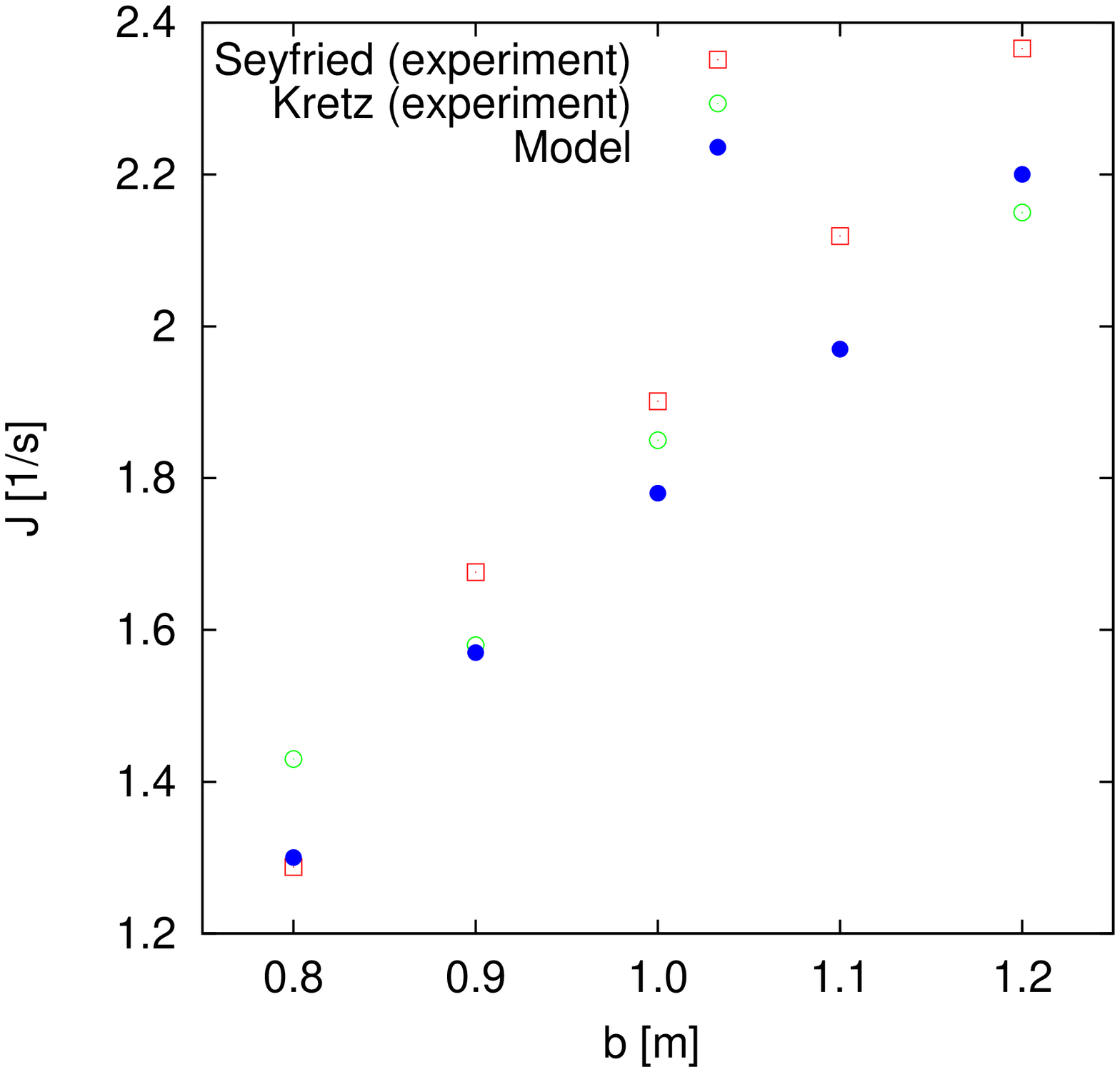}}
  \end{center}
 \caption{Left: The fundamental diagram in comparison with empirical
   data. For other values of the corridor's width
   ($1\;\mbox{m}\;\mbox{and}\; 4\;\mbox{m}$), the simulation results
   are also in good agreement with the empirical data. Right: Flow
   measurement with the modified CFM in comparison with empirical
   data.}
 \label{fig:fd2d-flow}
\end{figure}
The shape of the reproduced velocity-density relation is in good
agreement with the empirical data
\cite{Mori1987,Helbing2007,Oeding1963,Hankin1958}, see
Fig.~\ref{fig:fd2d-flow}).

Furthermore, the flow of $60$ pedestrians through the bottleneck as
described in~\cite{Seyfried2009b} was simulated. The width of the
bottleneck was  changed from $0.8\;$m to $1.2\;$m in steps of
$0.1\;$m (Fig.~\ref{fig:fd2d-flow}).

\begin{figure}[htp]
  \begin{center}
    \subfigure[Density in front of the entrance to the
    bottleneck]{\label{fig:rhovor}\includegraphics[scale=0.33, angle=270]{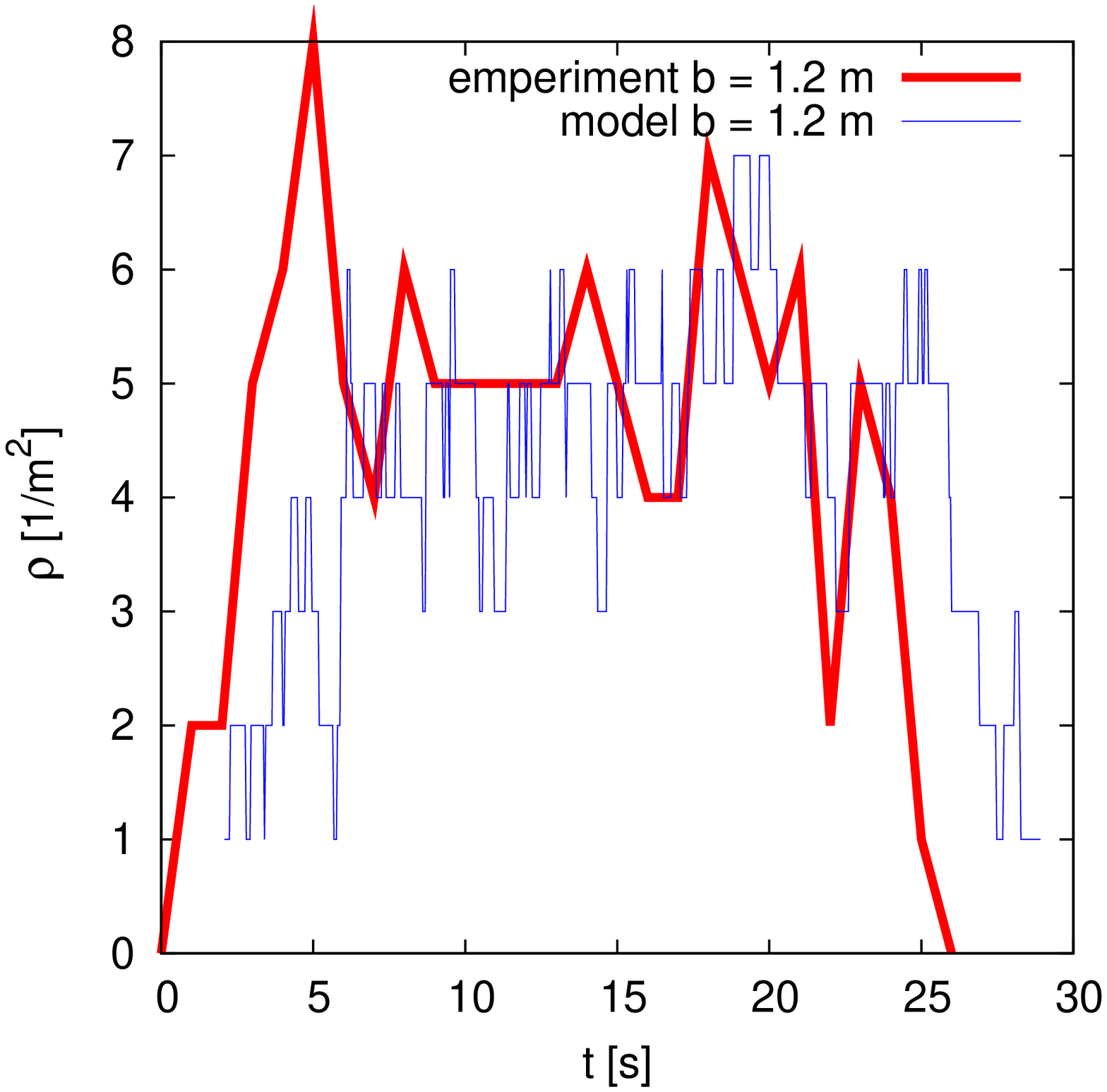}}
\subfigure[Density inside the
    bottleneck]{\label{fig:rhoin}\includegraphics[scale=0.33, angle=270 ]{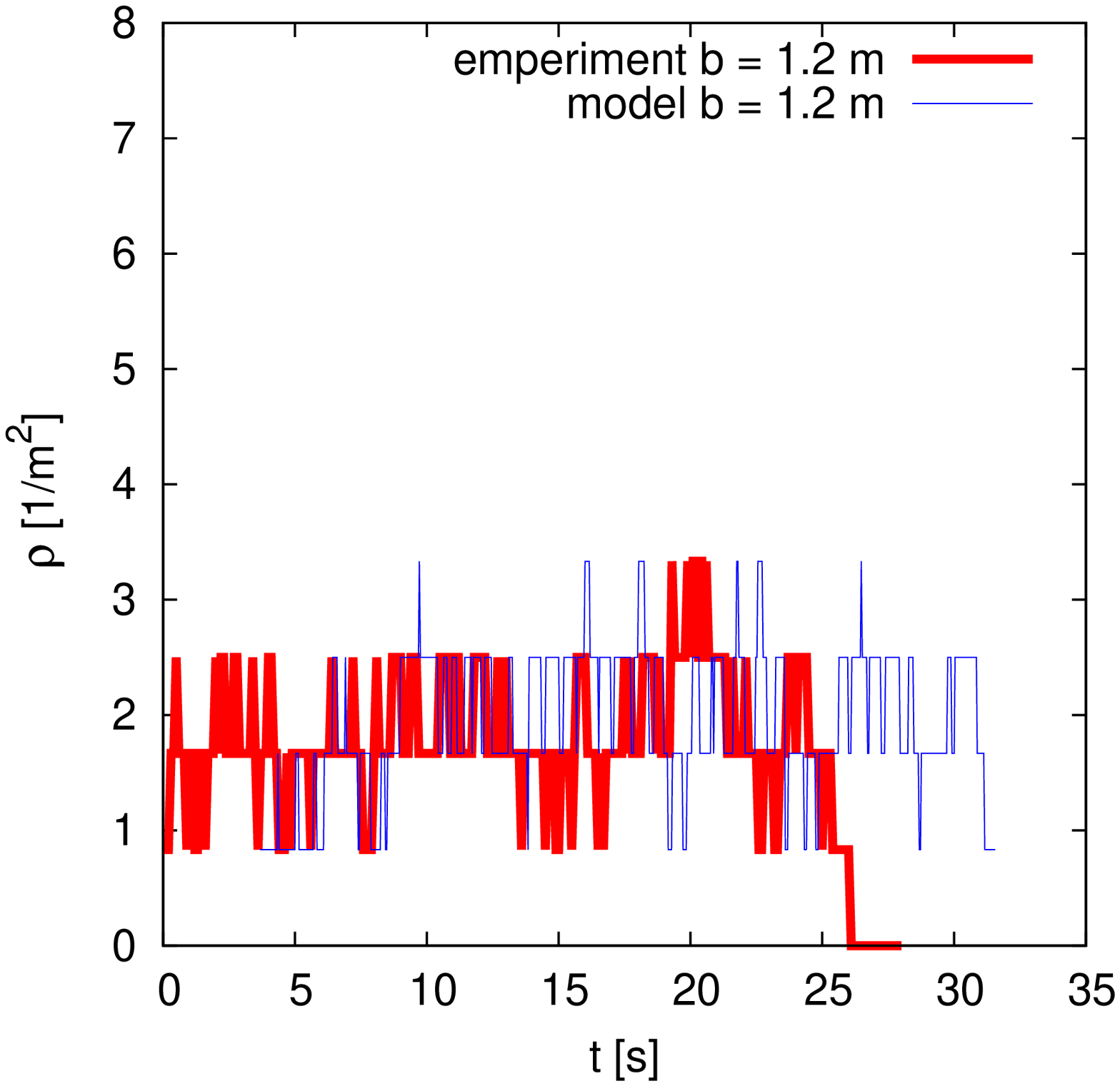}}
  \end{center}
  \caption{Density measurements: The simulation results (blue lines) are 
    in good agreement with the empirical data presented in
    \cite{Seyfried2009} and~\cite{Rupprecht2007}. The difference between
 the density in front and inside the bottleneck as well as the amplitude
 of the fluctuations are given correctly. The width of the
 bottleneck is $1.2\;\mbox{m}$. Also for other values of the width a
 good agreement between simulation results and empirical data is found.}
  \label{fig:rho}
\end{figure}

A third validation comes from measurements of density inside the
bottleneck as well as in front of the entrance to the bottleneck. The
density in front of the entrance to the bottleneck is presented in
Fig.~\ref{fig:rhovor}.  The results are in good agreement with the
experimental data in~\cite{Seyfried2009b}. Additionally, the measured
density values inside the bottleneck are in accordance with the
published empirical results in~\cite{Rupprecht2007}, see
Fig.~\ref{fig:rhoin}. One remarks that the density in front of the
bottleneck is much higher than the density in the bottleneck. This
difference reflects typical dynamics at bottlenecks, which is
reproduced by our model.

\section{Conclusions}
We have proposed modifications of a spatially continuous
force-based model~\cite{Yu2005} to describe quantitatively the
movement of pedestrians in 2D-space. Besides being a remedy for
numerical instabilities in CFM the modifications simplify the approach
of Yu et al.~\cite{Yu2005} since we can dispense with their extra ``collision
detection technique'' without deteriorating performance. The
implementation of the model is straightforward and does not use any
restrictions on the velocity. Simulation results show good agreement
with empirical data. Nevertheless, the model contains free parameters
that have to be tuned adequately to adapt the model to a given
scenario. Further improvement of the model could be made by
including, for example, a density dependent repulsive force.

\section*{Acknowledgement}
The authors are grateful to the Deutsche Forschungsgemeinschaft
(DFG) for funding this project under Grant-Nr.: SE 1789/1-1.

\end{document}